\documentclass{osa-articleb}
\journal{osac}

\newcommand\ee{\mathrm{e}}
\newcommand\cc{\mathrm{c}}
\newcommand\dd{\mathrm{\;d}}
\newcommand\nm{\mathrm{\;nm}}
\newcommand\TE{\mathrm{TE}}
\newcommand\TM{\mathrm{TM}}

\newcommand\ii{\mathrm{i}}
\newcommand\omegap{\omega_\mathrm{p}}

\renewcommand\Re{\mathop{\rm Re}\limits}
\renewcommand\Im{\mathop{\rm Im}\limits}
\newcommand\lt\left
\newcommand\rt\right

\newcommand\mat[1]{\mathbf{#1}}

\begin{document}

\title{Topologically protected bound states in~the~continuum and strong phase resonances in~integrated Gires--Tournois interferometer}

\author{Dmitry A. Bykov\authormark{1,2,\dag}, Evgeni A. Bezus\authormark{1,2,\dag}, \\ Leonid L. Doskolovich\authormark{1,2}}

\address{\authormark{1}Image Processing Systems Institute --- Branch of the Federal Scientific Research Centre ``Crystallography and Photonics'' of Russian Academy of Sciences, 151 Molodogvardeyskaya st., Samara 443001, Russia}
\address{\authormark{2}Samara National Research University, 34 Moskovskoye shosse, Samara 443086, Russia}

\address{\authormark{*}Corresponding author: evgeni.bezus@gmail.com}

\address{\authormark{$\dag$}These authors contributed equally to this work.}

\begin{abstract}
Photonic bound states in the continuum (BICs) are eigenmodes with an infinite lifetime, which coexist with a continuous spectrum of radiating waves.
BICs are not only of great theoretical interest, but have a wide range of practical applications, e.g.\ in the design of optical resonators.
Here, we study this phenomenon in a new integrated nanophotonic element consisting of a single dielectric ridge terminating an abruptly ended slab waveguide.
This structure can be considered as an on-chip analogue of the Gires--Tournois interferometer (GTI).
We demonstrate that, in contrast to the conventional GTI, the proposed integrated structure supports high-Q phase resonances and ``conditional'' BICs.
We develop a simple but extremely accurate coupled-wave model, which clarifies the physics of the BIC formation and enables predicting their locations.
We show that the studied BICs are topologically protected and, for the first time, investigate the strong phase resonance effect, which occurs when two BICs with opposite topological charges annihilate.
\end{abstract}

\section{Introduction}
Resonances are a key concept in photonics since they are behind many intriguing optical effects arising in various photonic structures.
Resonances occur when an eigenmode of the structure is excited,
which results in pronounced peaks in the transmission and reflection spectra, and in the local field enhancement, leading to extraordinary magneto-optical, nonlinear, and other optical effects.
This makes resonant structures an indispensable building block for a wide range of photonic devices.
\emph{High-Q} resonators are particularly important in the design of lasers, filters, and sensors.

In the last decade, a fascinating phenomenon allowing one to engineer resonances with an arbitrarily high quality factor drew a lot of interest in photonics~\cite{Marinica, Sadreev1, HsuReview, FanOpt}.
This phenomenon, referred to as bound states in the continuum (BICs), was first predicted for an electronic system 
by von~Neumann and Wigner in 1929~\cite{NW}.
In photonics, bound state in the continuum is a non-radiating eigenmode of a structure having open scattering channels.
The leakage of the BIC energy into these channels can be prevented by means of different mechanisms~\cite{HsuReview}, including 
  symmetry protection~\cite{Fan:2002:prb, Shipman, my:Bykov:2019}, 
  interaction of several resonators~\cite{Marinica, Ndangali}, 
  and interference of several resonances in the same cavity~\cite{FW, Lepetit, Bulgakov:2018:pra, my:Bezus:2018:pr, Bulgakov:2018:josab, my:Bykov:2019}.
The BICs have infinite quality factor, however, a slight deviation from a BIC allows one to obtain resonances with extremely high Q-factors.

Different photonic structures were shown to support BICs~\cite{HsuReview}.
Most papers study BICs in periodic structures, in particular, 
in photonic crystal slabs~\cite{Hsu, Shipman, Sadrieva2, Bulgakov:2018:pra, Marinica, Blanchard},
guided-mode resonant gratings~\cite{Bulgakov:2018:josab, my:Bykov:2019}, 
interfaces of photonic crystals~\cite{Hsu2},
and
infinite arrays of dielectric rods or spheres~\cite{Sadreev2, Blanchard, Yuan, Bulgakov:2017:prl}.
In all these structures, the open scattering channels are the ``free-space'' diffraction orders.
In papers~\cite{Sadreev1,A1,A2,A3,Zou, my:Bezus:2018:pr}, a different class of structures was studied with the scattering channels being the modes of photonic crystal waveguides~\cite{Sadreev1,A1,A2,A3}, 
slab waveguides~\cite{Zou, my:Bezus:2018:pr}, 
or rectangular microwave waveguides~\cite{Lepetit}.

It was recently shown that BICs can be endowed with a topological invariant --- topological charge, which takes integer values and is conserved upon the variation of the parameters of the structure~\cite{Zhen:2014:prl}.
The BICs having non-zero topological charge are robust to these variations.
Besides, the topological charge conservation law determines the possible interaction scenarios of several BICs~\cite{Zhen:2014:prl, Bulgakov:2017:pra, Bulgakov:2017:prl}.

Despite the large number of works on BICs published in recent years, the BICs associated with phase-only resonances have not yet been investigated.
At the same time, such a phenomenon would be of great interest for pulse dispersion engineering and phase encoding using nanophotonic structures.

In this paper, we propose a new integrated photonic structure --- an integrated Gires--Tournois interferometer (GTI) consisting of a single dielectric ridge terminating an abruptly ended slab waveguide.
We demonstrate that, in contrast to the conventional GTI~\cite{GT}, the proposed integrated analogue supports BICs and strong phase resonances.
We present an accurate analytical model describing the optical properties of the studied integrated structure and revealing the mechanism behind the BIC formation.
By introducing the topological charge of the BICs, we prove that they are robust.
Moreover, we show that the BICs supported by the considered structure are ``conditional'', i.\,e. they exist only when a certain condition is satisfied.
We formulate this (necessary and sufficient) condition in terms of three local coupling coefficients, which describe light scattering by the edges of the ridge.
When the condition is violated, the BICs having opposite topological charges group in pairs and annihilate with each other, leading to a strong phase resonance.


\section{Integrated Gires--Tournois interferometer}
\begin{figure}[ht]
	\centering\includegraphics{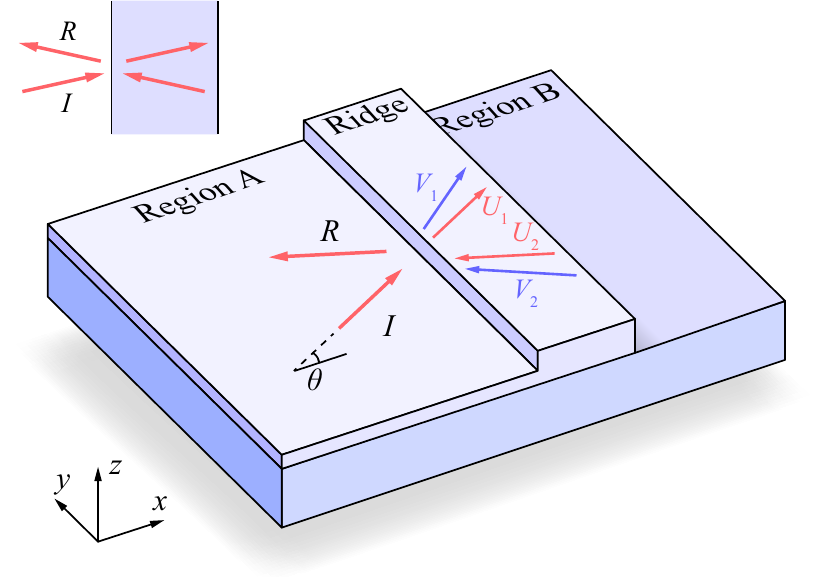}
	\caption{\label{fig:1}Geometry of the integrated Gires--Tournois interferometer (the structure is translation-invariant in the $y$-direction). The arrows depict the propagation directions of the slab waveguide modes inside the structure. Red arrows show TE-polarized modes; blue arrows show TM-polarized modes. The inset shows the conventional GTI.}
\end{figure}

First, let us recall the conventional Gires--Tournois interferometer (GTI), which dates back to 1964~\cite{GT}.
This interferometer (etalon) consists of a dielectric slab with the first interface being partially reflective, and the second one having unit reflectivity (see the inset to Fig.~\ref{fig:1}).
Similarly to the Fabry--P{\'e}rot interferometer, GTI exhibits resonant  optical properties due to multiple reflections between the interfaces of the slab.
If there is no absorption inside the GTI, the intensity of the reflected light is unity, whereas its phase changes in a resonant manner.
This makes GTI extremely important for various applications including dispersion compensation and compression of frequency modulated light pulses~\cite{GT, GT2, GT3}.

Let us consider an integrated analogue of the GTI shown in Fig.~\ref{fig:1}.
The proposed structure consists of a single dielectric ridge with the thickness $h_r$ and width $w$ terminating an abruptly ended slab waveguide with the thickness $h_a$ ($h_a < h_r$).
We restrict our consideration to the case when both the slab waveguides with the thicknesses $h_a$ and $h_r$ support only the fundamental TE- and TM-polarized guided modes.
However, the effects discussed below can also occur in thicker multimode waveguides.

We study the diffraction of the TE-polarized guided mode obliquely incident from region~A at the angle of incidence $\theta$ (Fig.~\ref{fig:1}).
We consider the monochromatic case with time dependence of $\exp(-\ii \omega t)$, where $\omega = 2\pi \cc/\lambda$ is the light angular frequency and $\lambda$ is the free-space wavelength.
The incident wave has a plane wavefront with the following $x$-$y$ dependence: $\exp(\ii k_x x + \ii k_y y)$, where 
\begin{equation}
\label{kxky}
k_x = k_0 n_{\mathrm{TE, inc}} \cos\theta,\;\;\; k_y = k_0 n_{\mathrm{TE, inc}} \sin\theta
\end{equation}
 are the wave vector components.
Here, $k_0=2\pi/\lambda$ is the wavenumber and $n_{\mathrm{TE, inc}}$ is the effective refractive index of the incident TE-polarized guided mode (Fig.~\ref{fig:1}).
In some works, such waves are referred to as semi-guided planar waves \cite{hammer1,hammer2}.

At small angles of incidence, the outgoing field includes not only the reflected TE- and TM-polarized guided modes, but also a continuum of propagating waves scattered out of the waveguide core layer.
However, at relatively large angles of incidence, all these ``parasitic'' non-guided waves become evanescent and do not carry energy.
We present a brief discussion of this scattering cancellation mechanism in Supplement 1; several integrated nanophotonic elements exploiting this phenomenon were recently proposed in~\cite{hammer1,hammer2, my:Bezus:2018:pr, tgpf, tgpd}.

Since there is no waveguide in region~B (see Fig.~\ref{fig:1}) and no out-of-plane scattering, all the energy is reflected at the second edge of the ridge and eventually returns to region~A.
Therefore, the structure works in the total internal reflection geometry.
Moreover, if $n_{\mathrm{TE, inc}}\sin\theta > n_{\mathrm{TM, inc}}$, where $n_{\mathrm{TM, inc}}$ is the effective refractive index of the TM-polarized guided mode in region A, the reflected field contains only the TE-polarized mode, which has the same intensity as the incident wave.

In this regime, the structure indeed turns out to be an integrated analogue of the Gires--Tournois interferometer:
the first interface of the ridge acts as a ``weak'' front mirror, whereas the second interface corresponds to a back mirror with unit reflectivity.
Thus, the ridge itself can be considered as the interferometer cavity.
There is, however, an important difference between the investigated on-chip structure and the conventional Gires--Tournois interferometer.
Indeed, at angles of incidence $\theta$ such that 
\begin{equation}
\label{thetarange}
n_{\mathrm{TM}} > n_{\mathrm{TE, inc}}\sin\theta > n_{\mathrm{TM, inc}},
\end{equation}
the outgoing field, as discussed above, consists of a single wave (the reflected TE mode), whereas in the ridge region, both the TE- and TM-polarized modes propagate.
Here, $n_{\rm TM}$ is the effective refractive index of the TM-polarized mode in the ridge region.
In the rest of the paper, we will focus on the angle of incidence range defined by Eq.~(\ref{thetarange}), and will show that in this angular range the structure possesses remarkable resonant optical properties.

To illustrate these properties, we consider an example with the following parameters: free-space wavelength $\lambda = 630$ nm; refractive indices of the superstrate, waveguide core layer, and substrate $n_{\mathrm{sup}} = 1$, $n_{\mathrm{inc}} = 3.32$ (GaP), and $n_{\mathrm{sub}} = 1.45$, respectively; waveguide thicknesses in region A and in the ridge region $h_a = 90\nm$ and $h_r = 110\nm$, respectively.
Let us emphasize that the used parameter values are not unique (no optimization of the structure was performed) and that the effects described below can be observed for a wide range of materials and waveguide thicknesses.

For the example under study, the slab waveguide in region A supports a TE-polarized mode with effective refractive index $n_{\mathrm{TE, inc}} = 2.681$ and a TM-polarized mode with effective refractive index $n_{\mathrm{TM, inc}} = 1.796$.
The slab waveguide with thickness $h_r$ corresponding to the ridge region supports TE- and TM-polarized modes with effective refractive indices $n_{\mathrm{TE}} = 2.819$ and $n_{\mathrm{TM}} = 2.187$, respectively.
At these parameters, the inequalities~(\ref{thetarange}) are satisfied at the angles of incidence $42.05^\circ < \theta < 54.64^\circ$.

\begin{figure*}[htb]
\hspace{-2.5cm}
\includegraphics{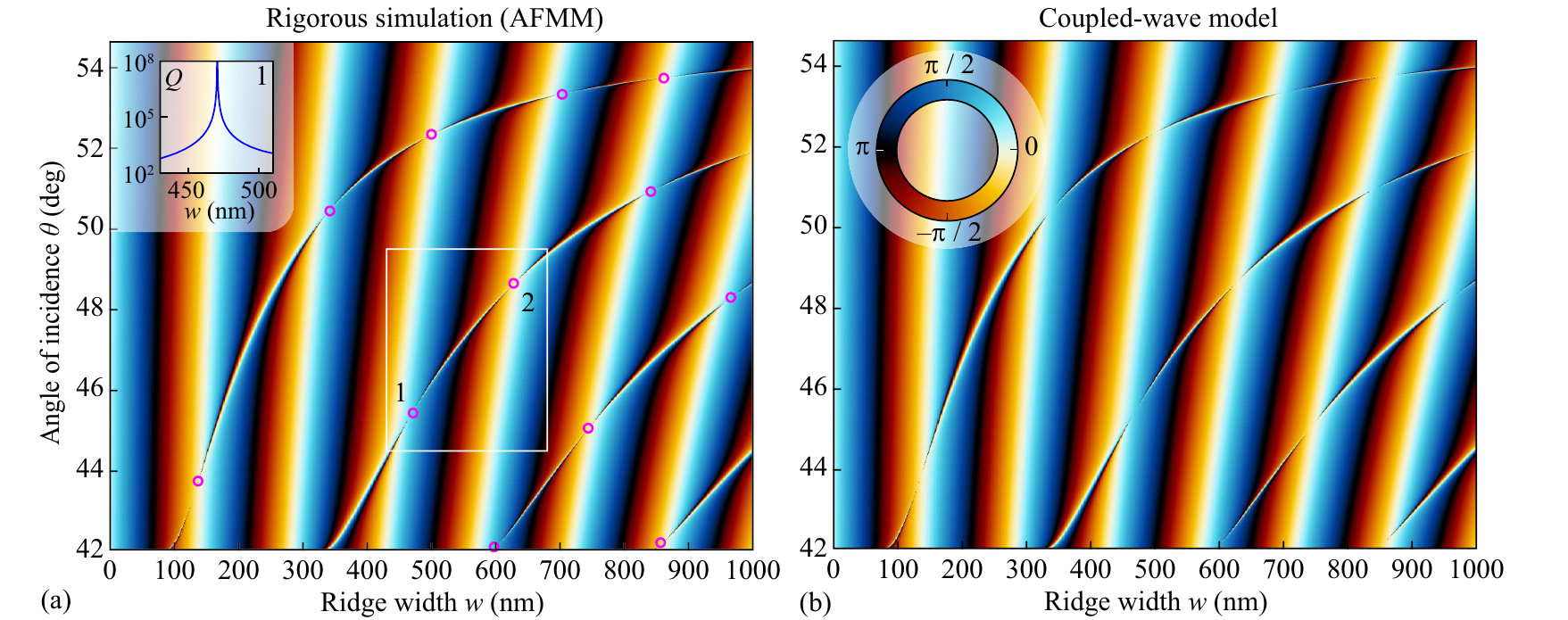}
\caption{\label{fig:2}Phase of the reflected TE-polarized mode ($\arg R$) vs.\ ridge width~$w$ and angle of incidence~$\theta$ calculated using the AFMM~(a) and using the proposed coupled-wave model~(b).
Magenta circles show the BIC positions predicted using the coupled-wave model.
The inset in~(a) shows the quality factor of the resonance near the BIC marked with ``1''.
The BICs ``1'' and ``2'' bounded by the white rectangle are investigated in detail in Section~\ref{sec:annih}.}
\end{figure*}

Since the reflected TE-polarized mode has a constant (unit) amplitude in the angular range of interest, it is the phase of the reflected radiation that has to be investigated.
The rigorously calculated dependency of the phase $\arg R$ on the ridge width $w$ and the angle of incidence $\theta$ is shown in Fig.~\ref{fig:2}(a).
The plot was calculated using an efficient in-house implementation of the aperiodic Fourier modal method (AFMM), an established numerical technique for solving Maxwell's equations in integrated optics problems~\cite{Silberstein:2001:josaa, Li:1996:josaa2, Hugonin:2005:josaa}.

Figure~\ref{fig:2}(a) demonstrates that along with the smooth fringes caused by Fabry--P\'erot resonances of TE-polarized guided modes in the ridge region, sharp resonant features are present in the phase spectrum.
The calculation of $\arg R$ in a wider angular range (not presented here) shows that the high-Q resonances occur only at the angles of incidence considered in Fig.~\ref{fig:2}.
As discussed above, at these angles, the reflected radiation contains only the TE-polarized guided mode, whereas in the ridge region, both the TE- and TM-polarized modes exist.
The fact that the high-Q resonance region is located between the cut-off angles of the TM modes outside and inside the ridge suggests that these sharp features arise due to the excitation of cross-polarized (quasi-TM) eigenmodes of the ridge.

However, the most interesting feature of Fig.~\ref{fig:2}(a) is that the width and, consequently, the quality factor of the phase resonances strongly vary along the dispersion curves.
At the points marked with magenta circles, the resonances vanish, i.\,e.\ their widths turn to zero and their quality factor diverges.
This is demonstrated by the rigorously calculated~\cite{Bykov:2015:co, my:bykov:2013:jlt, my:Bezus:2018:pr} Q-factor plot shown in the inset to Fig.~\ref{fig:2}(a), which indicates the presence of BICs in the considered structure.
In order to explain the BIC formation mechanism, in the following two sections we derive a simple but very accurate coupled-wave model, which proves the BIC existence and enables predicting their locations in the phase spectrum.

\section{Coupled-wave model}\label{sec:cwt}
In this section, we develop a coupled-wave model describing resonant optical properties of the considered structure.
The model is based on the fact that once we neglect the near-field effects, we can represent the field inside region A and inside the ridge region as a superposition of a small number of slab waveguide modes, which are coupled at the ridge edges.

As it was shown in the previous section, the structure exhibits resonant properties when it has only one scattering channel, namely, the reflected TE-polarized mode.
We denote the complex amplitude of the wave corresponding to this channel by $R$, whereas the amplitude of the incident wave will be denoted by $I$.
Inside the ridge region, there are both TE and TM slab waveguide modes with the amplitudes $U_{1,2}$ and $V_{1,2}$, respectively.
Here, the subscript ``1'' is used for the modes propagating from the first (left) edge of the ridge towards the second (right) one; the subscript ``2'' denotes the modes, which are reflected from the second edge and propagate towards the first one.
The propagation directions of these six waves are schematically shown in Fig.~\ref{fig:1}.

The considered modes are coupled at the edges of the ridge. 
The coupling at the first edge can be described by a $3\times3$ scattering matrix, which relates the complex amplitudes of the ``scattered'' modes ($R$, $U_1$, and $V_1$) with the amplitudes of the modes, which are incident on the first edge of the ridge ($I$, $U_2$, and $V_2$):
\begin{equation}
\label{S1}
\begin{bmatrix}
R \\ U_1 \\ V_1 
\end{bmatrix}
=
\underbrace{
\begin{bmatrix}
r & t_1 & t_2 \\
t_1 & r_1 & r_c \\ 
t_2 & r_c & r_2 
\end{bmatrix}
}_{\mat{S}_1}
\begin{bmatrix}
I \\ U_2 \\ V_2
\end{bmatrix}.
\end{equation}
Here, $r$ and $t$ denote reflection and transmission coefficients, respectively.
The subscripts denote the scattering channels being coupled: ``1'' and ``2'' correspond to the TE- and TM-polarized waves inside the ridge, whereas ``$c$'' corresponds to the cross-polarization coupling of the modes.
Let us note that these coefficients depend on the incidence angle $\theta$ and can be easily calculated using the AFMM~\cite{Silberstein:2001:josaa}.

Similarly, the coupling at the second edge of the ridge is described by a $2\times 2$  scattering matrix $\mat{S}_2$:
\begin{equation}
\label{S2}
\begin{bmatrix}
U_2 \ee^{-\ii \phi} \\ V_2 \ee^{-\ii \psi} 
\end{bmatrix}
=
\underbrace{
\begin{bmatrix}
q_1 & q_c \\ 
q_c & q_2 
\end{bmatrix}
}_{\mat{S_2}}
\begin{bmatrix}
U_1 \ee^{\ii \phi} \\ V_1 \ee^{\ii \psi} 
\end{bmatrix},
\end{equation}
where the subscripts have the same meaning as in Eq.~(\ref{S1}).
Since $U_{1,2}$ and $V_{1,2}$ denote the amplitudes of the modes at the first edge of the ridge, the scattering matrix $\mat{S}_2$ couples the amplitudes of the waves multiplied by the exponents describing the phase change due to the propagation of the modes between the ridge edges.
The phases $\phi$ and $\psi$ can be obtained by multiplying the wave vector components of the modes by the ridge width $w$:
\begin{equation}
\label{phipsidef}
 \phi = k_x^{\TE} w, \;\;\; \psi = k_x^{\TM} w,
\end{equation}
where
\begin{equation}
\label{kdef}
\left(k_x^{\TE}\right)^2 + k_y^2 = \left(\frac{\omega}{\cc} n_{\TE}\right)^2,
\;\;
\left(k_x^{\TM}\right)^2 + k_y^2 = \left(\frac{\omega}{\cc} n_{\TM}\right)^2,
\end{equation}
and $k_y$ is defined in Eq.~(\ref{kxky}).

Equations~(\ref{S1}) and~(\ref{S2}) give the system of five coupled-wave equations describing the optical properties of the considered structure.
When the amplitude of the incident wave is zero ($I = 0$), this system of linear equations becomes homogeneous, having a non-trivial solution only when the determinant of the matrix of the system
\begin{equation}
\label{D}
\begin{aligned}
\mathcal{D} &= \lt(r_1 q_1 -\ee^{-2\ii \phi}\rt)\lt(r_2 q_2 - \ee^{-2\ii \psi}\rt) 
\\
&\;\;\;\;- 2 q_c r_c \ee^{-\ii \phi}\ee^{-\ii \psi} 
 - \lt( q_1 q_2 - q_c^2 \rt) r_c^2 - q_c^2 r_1 r_2
\end{aligned}
\end{equation}
vanishes.
This solution describes the eigenmodes of the structure with the equation 
\begin{equation}
\label{Dzero}
\mathcal D = 0
\end{equation}
being its dispersion relation.

Let us assume that the incident wave has unit amplitude ($I = 1$). 
In this case, we can solve Eqs.~(\ref{S1}) and~(\ref{S2}) for~$R$, which gives us the complex reflection coefficient in the form of a fraction:
\begin{equation}
\label{mainR1}
R = \frac{\mathcal{N}}{\mathcal{D}},
\end{equation}
where the denominator is given by Eq.~(\ref{D}), whereas the numerator reads as
\begin{equation}
\label{N}
\begin{aligned}
\mathcal{N} = r \mathcal{D} &+
q_1 t_1^2 \ee^{-2\ii \psi} + 2 q_c t_1 t_2  \ee^{-\ii \phi}\ee^{-\ii \psi} + 
 q_2 t_2^2 \ee^{-2\ii \phi} \\
&- \lt(q_1 q_2 - q_c^2\rt) \lt(r_2 t_1^2 -2 r_c t_1 t_2 + r_1 t_2^2\rt).
\end{aligned}
\end{equation}

Despite the fact that the obtained expressions~(\ref{D}) and~(\ref{N}) are quite complicated, it is possible to carry out their detailed and fruitful analysis.
This analysis is based on enforcing the energy conservation law, which ensures the unitarity of the scattering matrices $\mat{S}_1$ and $\mat{S}_2$.
In Supplement~1, we derive several simple relations, which follow from the unitarity property and allow us to rewrite the reflection coefficient~(\ref{mainR1}) in the following form:
\begin{equation}
\label{mainR2}
R = \ee^{2\ii\phi}\ee^{2\ii\psi} d_1\, d_2\frac{\mathcal{D}^*}{\mathcal{D}}.
\end{equation}
Here, $\mathcal{D}^*$ is the complex conjugate of $\mathcal{D}$ and
 $d_i=\det\mat{S}_i$, $i=1,2$ denote the scattering matrix determinants, which have unit magnitudes since the matrices $\mat{S}_1$ and $\mat{S}_2$ are unitary.
The presented form of the reflection coefficient makes it obvious that $|R| = 1$.

Figure~\ref{fig:1}(b) shows the phase of the reflected light ($\arg R$) calculated using Eq.~(\ref{mainR2}).
By comparing Figs.~\ref{fig:1}(a) and~\ref{fig:1}(b), we see that the developed coupled-wave model is in excellent agreement with the full-wave simulation results.
In particular, the presented model reproduces the sharp phase resonances with varying width, which are present in the rigorously calculated spectrum of Fig.~\ref{fig:1}(a).
In the next section, we further analyze Eq.~(\ref{mainR1}) and show that the proposed model proves the BIC existence and allows one to calculate their locations.

\section{Bound states in the continuum}\label{sec:bic}
Bound states in the continuum (BICs) are the eigenmodes of the structure, hence, they must satisfy Eq.~(\ref{Dzero}). 
At the same time, BICs have infinite quality factor, therefore, Eq.~(\ref{Dzero}) holds at a real frequency $\omega$.
Since~$\mathcal{D}$ is the denominator of the reflection coefficient~$R$ [see Eq.~(\ref{mainR1})], in the case of a BIC, the zero in the denominator~$\mathcal{D}$ should be compensated by a zero in the numerator~$\mathcal{N}$~\cite{Blanchard, my:Bezus:2018:pr, my:Bykov:2019}.
Therefore, the BICs can be found as real-$w$ and real-$\theta$ solutions of the following system of equations:
\begin{equation}
\label{NDsys}
\left\{
\begin{aligned}
\mathcal{D} = 0, \\
\mathcal{N} = 0.
\end{aligned}
\right.
\end{equation}

This system can be easily solved with respect to the exponents $\ee^{-2\ii\phi}$ and $\ee^{-2\ii\psi}$.
Taking into account the unitarity of the scattering matrices, we can write the solution as
\begin{equation}
\label{eqmain2}
\begin{aligned}
	\ee^{-2\ii \phi} &= 
	d_1\frac{q_c^2}{2 q_2 t_2^2}\cdot
	\lt[2\frac{|t_2|^2}{|q_c|^2}  - \lt( |t_1|^2 + |t_2|^2 \rt) \pm \sqrt{\xi}\rt],\\
	\ee^{-2\ii \psi} &= 
	d_1\frac{q_c^2}{2 q_1 t_1^2}\cdot
	\lt[2\frac{|t_1|^2}{|q_c|^2} - \lt( |t_1|^2 + |t_2|^2 \rt) \pm  \sqrt{\xi}\rt].
\end{aligned}
\end{equation}
Here, the same sign of the square root has to be used in both equations.
The radicand~$\xi$ is given by the following real number:
\begin{equation}
\label{eqxi}
\xi = \left(|t_1|^2+|t_2|^2\right)^2 - \frac{4 |t_1|^2 |t_2|^2}{|q_c|^2},
\end{equation}
The detailed derivation of Eqs.~(\ref{eqmain2}) and~(\ref{eqxi}) is presented in Supplement~1.

According to Eq.~(\ref{phipsidef}), the phases $\phi$ and $\psi$ have to be real.
Therefore, the left-hand sides and, hence, the right-hand sides in Eqs.~(\ref{eqmain2}) must lie on the unit circle in the complex plane.
In Supplement~1, we prove that it is true once the argument $\xi$ of the square roots in Eqs.~(\ref{eqmain2}) is negative. This condition can be rewritten as a constraint on the cross-polarization reflection coefficient $q_c$:
\begin{equation}
\label{biccond}
|q_c| \le \frac{2 |t_1| |t_2|}{|t_1|^2+|t_2|^2}.
\end{equation}
According to this inequality, the BICs exist when the cross-polarization reflection coefficient $q_c$ is relatively small.
In particular, BICs always exist when no cross-polarization coupling occurs at the second edge of the ridge (at $q_c = 0$). 

Once inequality~(\ref{biccond}) is satisfied, the phases $\phi$ and $\psi$ providing the BIC condition are real and can be found by taking the argument of Eq.~(\ref{eqmain2}). 
Moreover, we can find real $\theta$ and $w$ providing these values of $\phi$ and $\psi$, and, hence, resulting in a BIC.
The closed-form expressions for the BIC positions in the $w$--$\theta$ parameter space
can be found by solving Eqs.~(\ref{phipsidef}) and~(\ref{kdef}):
\begin{equation}
\label{bicpos}
\begin{aligned}
w = \frac{\cc}{\omega}&\sqrt\frac{\phi^2-\psi^2}{n_\TE^2-n_\TM^2},\\
\theta = \arcsin&\sqrt\frac{n_\TM^2 \phi^2-n_\TE^2 \psi^2}{n_{\rm TE, inc}^2(\phi^2-\psi^2)}.
\end{aligned}
\end{equation}
The BIC positions calculated using Eqs.~(\ref{bicpos}) and shown in Fig.~\ref{fig:2}(a) are in perfect agreement with the features in the rigorously calculated phase of the reflected wave.
Let us note that the positions of different BICs in Fig.~\ref{fig:2}(a) were calculated by adding different integer multiples of $2\pi$ to the phases $\phi$ and $\psi$ in Eqs.~(\ref{bicpos}).


\section{Topological charge of the BICs}\label{sec:topo}
In this section, we demonstrate that the BICs in the considered structure are topologically protected.
We do this by showing that to each BIC, a non-zero integer topological charge can be assigned.
Usually, topological charge is defined as a curvilinear integral of some quantity calculated along a curve encircling a BIC in a certain parameter space.
Topological charge is introduced differently depending on the geometry of the structure. Approaches exploiting far-field polarization vector~\cite{Zhen:2014:prl, Bulgakov:2017:pra} and argument of the quasimodal expansion coefficient~\cite{Bulgakov:2017:prl} were recently proposed.

For the considered structure, we propose the following quantity to define the topological charge:
\begin{equation}
\label{P}
\mathcal{P} = d \ee^{\ii \phi} \left( q_c t_1^* \ee^{\ii \psi} - q_1 t_2^* \ee^{\ii \phi}  \right) - t_2,
\end{equation}
which can be interpreted as the coupling coefficient between the incident light and the eigenmode of the integrated Gires--Tournois interferometer (see Supplement~1).
To define the topological charge, we take the contour integral of the gradient of the argument of $\mathcal{P}$:
\begin{equation}
\label{Charge}
C = \frac1{2\pi}\oint_\gamma{\dd \arg\mathcal{P}(w, \theta)},
\end{equation}
where $\dd \arg\mathcal{P}(w, \theta) = \frac{\partial\arg\mathcal{P}}{\partial w} \dd{w} + \frac{\partial\arg\mathcal{P}}{\partial\theta} \dd{\theta}$.
We note here that the phases $\phi$ and $\psi$ in $\mathcal{P}$ depend on both $\theta$ and $w$, whereas the coupling coefficients $q_{1,c}$ and $t_{1,2}$ depend solely on $\theta$.

When the integration path $\gamma$ encircles a BIC, Eq.~(\ref{Charge}) defines its topological charge.
Moreover, when the topological charge is non-zero, there should be a point inside the integration contour where the phase of $\mathcal{P}$ is undefined. Therefore, at this point $\mathcal{P} = 0$. 
In Supplement~1, we prove that at this very point of the parameter space the considered structure supports a BIC.

\begin{figure}
\centering\includegraphics{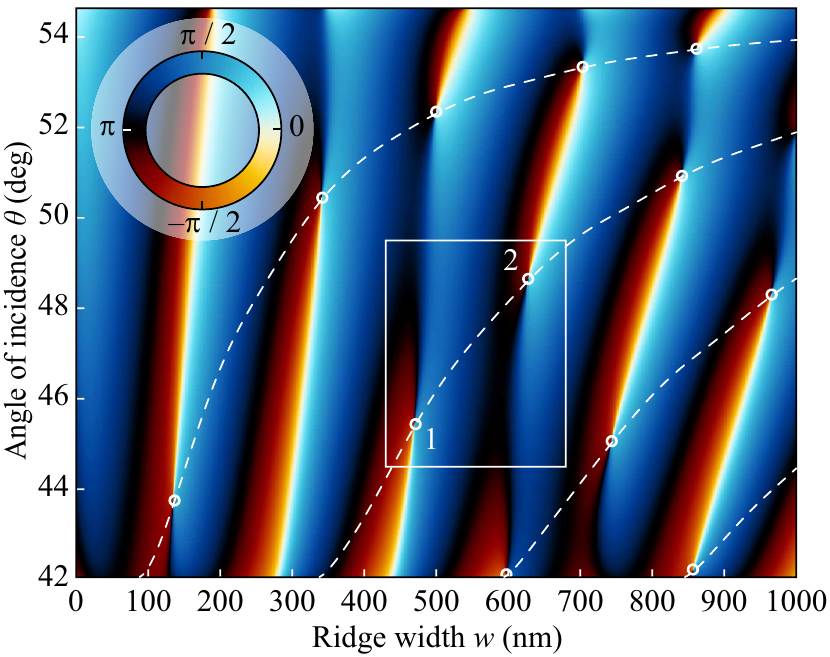}
\caption{\label{fig:5}Phase of $\arg \mathcal{P}$ vs.\ ridge width~$w$ and angle of incidence~$\theta$. As a guide to the eyes, the dashed lines show the dispersion of the quasi-TM modes of the ridge. White circles show the BIC positions predicted by the coupled-wave model.}
\end{figure}

Figure~\ref{fig:5} shows $\arg\mathcal{P}$ calculated in the considered $w$--$\theta$ parameter space.
The BICs are marked with white circles.
One can see the phase singularities appearing at these points.
Hence, if we move along a contour encircling a BIC, the phase will not return to its initial position but will differ by an integer multiple of $2\pi$.
For example, in the case of the BIC marked as ``1'' in Fig.~\ref{fig:5} the phase increases by $2\pi$ when encircling the BIC counter-clockwise, therefore, its topological charge is $+1$.
The phase around the BIC marked as ``2'', however, changes in the opposite direction, resulting in the topological charge $-1$.

Small perturbations of the parameters of the structure result in small changes in the quantity $\mathcal{P}$.
The integer value of the topological charge $C$, however, can only change in a discrete manner.
Therefore, small perturbations will not change the topological charge, and, therefore, the BIC will not disappear but move in the parameter space. Such BICs are called robust or topologically protected.
The calculations based on Eq.~(\ref{Charge}) show that all the BICs investigated in the current paper have non-zero topological charge and are, hence, robust.

Another important application of the topological charge is the investigation of the interaction of several BICs.
To do this, let us choose the integration contour $\gamma$ in such a way that it encircles several BICs. 
In this case, the value of $C$ in Eq.~(\ref{Charge}) will be the sum of the topological charges of the encircled BIC. 
If the BICs interact in some manner, after the interaction the sum of the topological charges is conserved.
One scenario for such interaction is the annihilation of the BICs having opposite charges, which will be discussed in the next section. 

\section{Annihilation of the BICs and strong phase resonances}\label{sec:annih}
As we have shown in the previous section, adjacent BICs in the considered structure have
 opposite topological charges ($-1$ and $+1$). 
The total topological charge of the two BICs is zero.
Therefore, according to the charge conservation law, the coalescence of these BICs may result in their annihilation, after which no BICs remain resulting in the same total charge equal to zero.
In this section, we demonstrate this effect. 

\begin{figure}
\centering\includegraphics{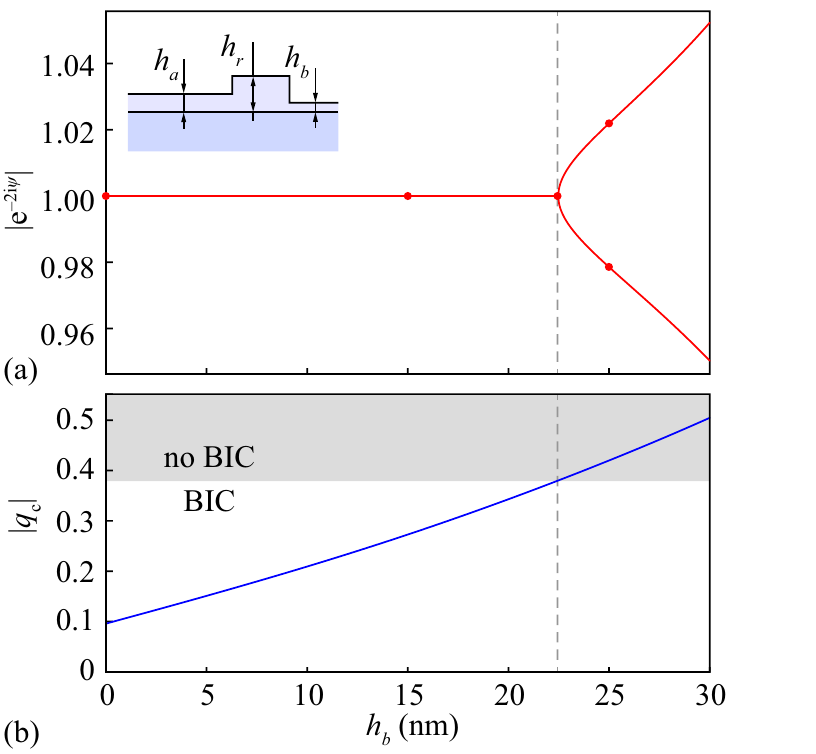}
\caption{\label{fig:3}Magnitude of the exponent $\ee^{-2\ii\psi}$ calculated using Eq.~(\ref{eqmain2})~(a) and of the cross--polarization coupling coefficient $q_c$~(b) vs.\ thickness $h_b$ at $\theta = 47.2^\circ$. The inset shows the geometry of the considered structure.}
\end{figure}

To ``move'' the BICs in the parameter space, we introduce a non-zero thickness $h_b$ of the waveguide layer in region~B (see the inset to Fig.~\ref{fig:3}).
We will assume that $h_b$ is small  enough, so that the scattering channels corresponding to the transmitted modes in region B remain closed.
In this case, the coupled-wave model of Section~\ref{sec:cwt} is still applicable.

\begin{figure*}
\hspace{-2.5cm}
\includegraphics{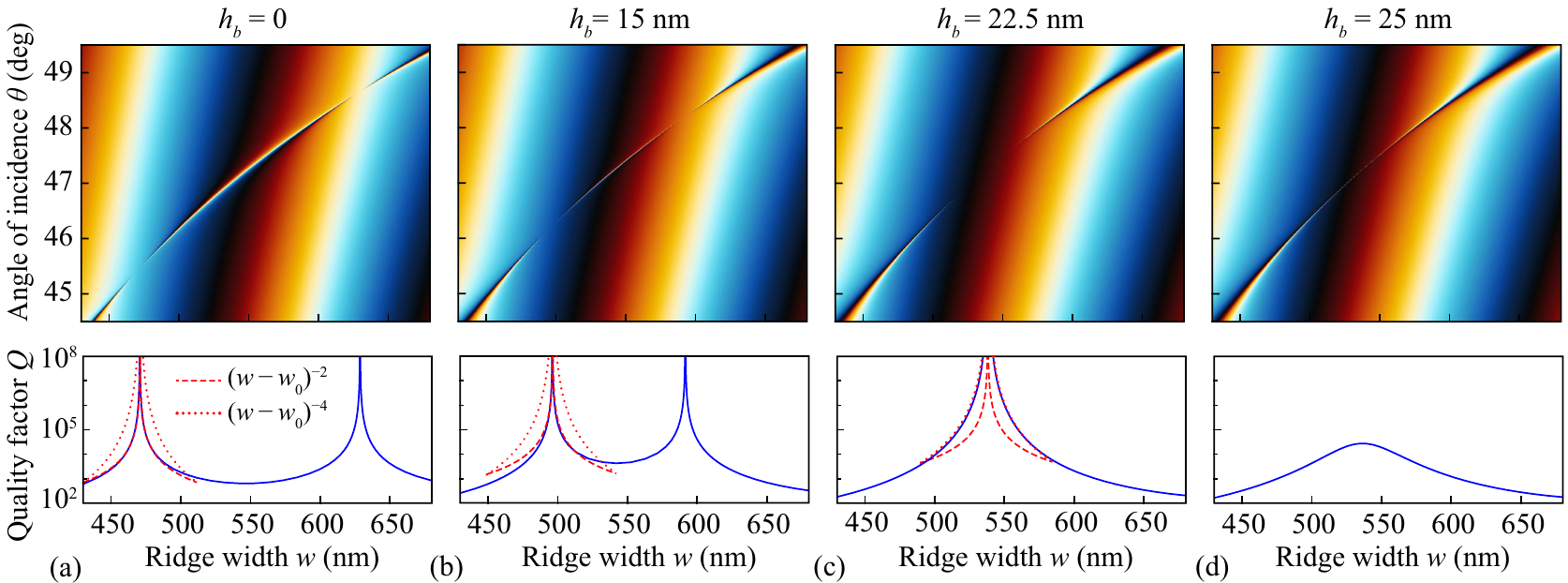}
\caption{\label{fig:4}Phase of the reflected wave $\arg R$ (upper panels) and the corresponding quality factor of the resonance (lower panels) at $h_b = 0$~(a), $h_b = 15\nm$~(b), $h_b = 22.5\nm$~(c), and $h_b = 25\nm$~(d). Red lines show the fitting of the curves $(w - w_0)^{-2}$ (dashed lines) and $(w - w_0)^{-4}$ (dotted lines) to the calculated quality factor plots. The color map of Fig.~\ref{fig:2} is utilized.}
\end{figure*}

Let us focus on the two BICs located in the region bounded by a white rectangle in Fig.~\ref{fig:2}(a).
In the upper panels of Fig.~\ref{fig:4}, we show the phase of the reflected wave calculated at different values of $h_b$ ranging from 0 to $25\nm$. 
The lower panels of Fig.~\ref{fig:4} show the corresponding rigorously calculated~\cite{Bykov:2015:co} quality factors of the modes.
Note that the upper panel of Fig.~\ref{fig:4}(a) is simply a magnified fragment of Fig.~\ref{fig:2}(a), where two BICs are present.
With an increase in $h_b$, the BICs approach each other as illustrated by Fig.~\ref{fig:4}(b).
Then, at $h_b = 22.5\nm$, the BICs coalesce, as it is shown in Fig.~\ref{fig:4}(c).
In this case, the topological protection of the BIC is lifted, i.\,e.\ the considered ``second-order'' BIC in Fig.~\ref{fig:4}(c) 
gets topological charge zero.
If we further increase $h_b$, the BIC will disappear; indeed, instead of a BIC, only a finite-Q resonance is seen in Fig.~\ref{fig:4}(d) corresponding to $h_b = 25\nm$.
This demonstrates the annihilation of the BICs governed by their topological properties.

Now, let us focus on the lower panels of Fig.~\ref{fig:4} and investigate the $w$-dependence of the quality factor near the value $w_0$ satisfying the BIC condition. 
To do this, we write the complex frequency of the mode $\omegap$ in the big~O notation:
$\Re \omegap = \omega+{\rm O}(w-w_0)$, $\Im\omegap = {\rm O}\lt((w-w_0)^2\rt)$, which follows from the causality condition~\cite{Blanchard}.
This expansion leads to a quadratic decay rate of the quality factor when moving away from the BIC: $Q = \Re \omegap/(-2\Im \omegap) \sim (w - w_0)^{-2}$.
Indeed, in the case of Figs.~\ref{fig:4}(a) and~\ref{fig:4}(b), the quadratic decay law $(w - w_0)^{-2}$ shown with dashed lines  provides a good fitting of the quality factor.
However, this is not the case in Fig.~\ref{fig:4}(c), where the two BICs coalesce and the decay rate becomes quartic: $Q \sim (w - w_0)^{-4}$ (see the dotted lines in Fig.~\ref{fig:4}).
In Refs.~\cite{Yuan:2017:pra, Bulgakov:2017:prl, Bulgakov:2017:pra, Yuan:2018:pra}, a similar phenomenon was predicted for different structures (periodic arrays of cylinders or spheres) and referred to as a strong resonance.
Therefore, in the case of a ``second-order'' BIC, the proposed integrated Gires--Tournois interferometer exhibits a strong \emph{phase} resonance, which  provides a slower decay of the quality factor when moving away from the BIC.
This effect is of practical importance for the design of high-Q resonators.

To gain a deeper understanding of the BIC annihilation mechanism, it is interesting to discuss how it is connected with the violation of the BIC condition~(\ref{biccond}) obtained in Section~\ref{sec:bic}.
Let us consider the dependence of the cross-polarization coupling coefficient $q_c$ on the thickness $h_b$ of the waveguide in region~B.
According to Fig.~\ref{fig:3}(b), an increase in $h_b$ results in an increase in $|q_c|$.
At $h_b = 22.5\nm$, the value of $|q_c|$ reaches the right-hand side of inequality~(\ref{biccond}), which then turns into an equality.
When $h_b$ exceeds $22.5\nm$, the BIC condition~(\ref{biccond}) is violated and the BIC ceases to exist, which agrees with the simulation results of Fig.~\ref{fig:4}(d).
Therefore, once the BIC condition~(\ref{biccond}) becomes violated, the BICs having opposite topological charges group in pairs and annihilate with each other.

In Section~\ref{sec:bic}, we also formulated the BIC condition in terms of the magnitudes of the exponents $\ee^{-2\ii\phi}$ and $\ee^{-2\ii\psi}$ that should have unit moduli in order for BICs to exist.
Figure~\ref{fig:3}(a) shows the magnitude of the exponent $\ee^{-2\ii\psi}$ calculated using Eqs.~(\ref{eqmain2}).
This magnitude equals unity when the BICs exist (at $h_b \leq 22.5\nm$),
but when the BIC condition is violated, the value of $|\ee^{-2\ii\psi}|$ is no longer unity. 
The same behavior can be observed for $|\ee^{-2\ii\phi}|$.
The two branches seen in Fig.~\ref{fig:3}(a) correspond to the positive and negative signs of the square root in Eqs.~(\ref{eqmain2}).
Thus, mathematically, the annihilation of the BICs is associated with the square-root anomaly in Eqs.~(\ref{eqmain2}).

\section{Conclusion}

In this work, we demonstrated that an integrated analogue of the Gires--Tournois interferometer (GTI), consisting of a single dielectric ridge terminating an abruptly ended slab waveguide, supports topologically protected bound states in the continuum (BICs).
We developed a  coupled-wave model accurately describing the resonant optical properties of the proposed structure.
In particular, simple closed-form expressions were obtained for the condition of the BIC existence and for the positions of the BICs in the considered parameter space. 
The coupled-wave model also allowed us to introduce the topological charge of the BICs, proving their robustness and limiting their possible interaction scenarios by the charge conservation law.

Detuning from the BIC condition allows one to obtain an all-pass filter exhibiting sharp phase resonances.
This makes the proposed structure promising for on-chip dispersion engineering, in particular, for phase equalization and pulse compression.
The existence of the so-called strong resonances in the considered structure relaxes the fabrication tolerances and makes the proposed integrated GTI an excellent candidate for creating high-Q optical resonators, which are widely used in filtering, sensing, and lasing.
The latter application may fruitfully exploit the fact that the proposed structure has only one open scattering channel.

We believe that the proposed concept of an integrated GTI supporting BICs and strong phase resonances can be extended to other on-chip platforms, in particular, Bloch surface electromagnetic waves propagating along the interfaces of photonic crystals~\cite{Vinogradov:2015:pu}.
We also expect that the demonstrated charge-conserving interaction of the BICs including their coalescence and annihilation can be implemented not only by changing the geometry of the structure, but also through an external stimulus of magneto-optical, electro-optical, or nonlinear nature.

\section*{Funding}
This work was funded by Russian Foundation for Basic Research (project nos. 18-37-20038 and 16-29-11683;
coupled-wave model) and by Ministry of Science and Higher Education of the Russian Federation (State assignment
to the FSRC ``Crystallography and Photonics'' RAS; numerical simulations).

\vspace{7.5em}

\clearpage

\begin{center}
\Large Supplementary materials / Appendices
\end{center}

\appendix

\section{Scattering channels}
Let us briefly discuss the scattering channels of the investigated integrated structure.
As it is mentioned in the main text of the paper, in the general case, the scattered field contains the reflected TE- and TM-polarized guided modes of the slab waveguide in region A as well as a continuum of ``parasitic'' non-guided waves scattered out of the waveguide.
However, let us show that this is not the case at relatively large angles of incidence $\theta$.

From the dispersion relation of a slab waveguide~\cite{lifante}, it follows that $n_{\mathrm{TE,inc}} > n_{\mathrm{TM,inc}} > \max(n_{\mathrm{sub}}, n_{\mathrm{sup}})$, where $n_{\mathrm{TE,inc}}$ and $n_{\mathrm{TM,inc}}$ are the effective refractive indices of the TE- and TM-polarized guided modes of the slab waveguide in region~A, and $n_{\mathrm{sub}}$ and $n_{\mathrm{sup}}$ are the refractive indices of the substrate and the superstrate, respectively.
A similar inequality can be written for the slab waveguide with the thickness $h_r$: $n_{\mathrm{TE}} > n_{\mathrm{TM}} > \max(n_{\mathrm{sub}}, n_{\mathrm{sup}})$, where $n_{\mathrm{TE}}$ and $n_{\mathrm{TM}}$ are the respective effective refractive indices of the TE- and TM-polarized guided modes in the ridge region.
Assuming that $h_r > h_a$, we obtain the inequalities $n_{\mathrm{TE}} > n_{\mathrm{TE,inc}}$ and $n_{\mathrm{TM}} > n_{\mathrm{TM,inc}}$~\cite{lifante}.
Finally, for the example considered in this work, all these inequalities can be combined into the following inequality: 
\begin{equation}
\label{ineqn}
n_{\mathrm{TE}} > n_{\mathrm{TE,inc}} > n_{\mathrm{TM}} > n_{\mathrm{TM,inc}} > n_{\mathrm{sub}} > n_{\mathrm{sup}}.
\end{equation}

Since all the interfaces of the structure are parallel to the $y$ axis, the wave vector component $k_y = k_0 n_{\mathrm{TE, inc}} \sin\theta$ is conserved for all the waves constituting the solution of the diffraction problem.
Here, $k_0 = 2\pi/\lambda$ is the wavenumber and $\lambda$ is the free-space wavelength.
From Eq.~(\ref{ineqn}), it follows that there exist angles of incidence $\theta > \arcsin(n_{\mathrm{sub}} / n_{\mathrm{TE,inc}})$, at which all the non-guided waves scattered out of the waveguide to the superstrate and substrate become evanescent.
A comprehensive description of this scattering cancellation mechanism is given in ~\cite{hammer1, hammer2, my:Bezus:2018:pr}.

If the angle of incidence is further increased and satisfies the inequality $\theta > \arcsin(n_{\mathrm{TM,inc}} / n_{\mathrm{TE,inc}})$, the reflected TM-polarized mode in region A also becomes evanescent. Thus, in this regime, a single scattering channel remains open in the considered integrated nanophotonic structure, namely, the reflected TE-polarized guided mode.

\section{Consequences of the unitarity of symmetric scattering matrices}\label{sec:A}

Let us consider a symmetric unitary matrix 
$$
\mat{S}_2 = 
\begin{bmatrix}
q_1 & q_c\\ 
q_c & q_2
\end{bmatrix}.
$$
Since the rows of a unitary matrix are normalized, we obtain  
\begin{equation}
\label{m2simple}
|q_1|^2=|q_2|^2=1-|q_c|^2.
\end{equation}
Moreover, we can equate the corresponding elements of the matrix inverse and of its conjugate transpose ($S^{-1} = S^\dagger$).
By doing so, we obtain the following useful relations:
\begin{equation}
\label{m2eqs}
\frac{q_2}{d_2} = q_1^*, \;\;
-\frac{q_c}{d_2} = q_c^*, \;\;
\frac{q_1}{d_2} = q_2^*,
\end{equation}
where $d_2 = \det{\mat{S}_2} = q_1 q_2 - q_c^2$. 
From the second equation, we obtain
\begin{equation}
\label{qprops}
\frac{q_1 q_2}{q_c^2} = 1-|q_c|^{-2},
\end{equation}
hence $q_1 q_2/q_c^2$ is a real number. Moreover, this number is non-positive since $|q_c|$ does not exceed unity.

Now, let us consider a $3\times 3$ unitary matrix 
$$
\mat{S}_1 = 
\begin{bmatrix}
r & t_1 & t_2 \\
t_1 & r_1 & r_c \\ 
t_2 & r_c & r_2 
\end{bmatrix}.
$$
By equating the matrix inverse to its conjugate transpose, we obtain the following simple relations:
\begin{equation}
\label{m3eqs}
\begin{aligned}
\frac{r_1 r_2 - r_c^2}{d_1} &= r^*, \;\;&
\frac{r_c t_2 - r_2 t_1}{d_1} &= t_1^*, \;\;&
\frac{r_c t_1 - r_1 t_2}{d_1} &= t_2^*, \\
\frac{r r_2 - t_2^2}{d_1} &= r_1^*, \;\;&
\frac{t_1 t_2-r r_c}{d_1} &= r_c^*,\;\;&
\frac{r r_1 - t_1^2}{d_1} &= r_2^*,
\end{aligned}
\end{equation}
where $d_1 = \det \mat{S}_1$. 
This determinant can be written as 
$d_1 = r (r_1 r_2 - r_c^2) + 2 (r_c t_2 - r_2 t_1) t_1 + (r_2 t_1^2 - r_1 t_2^2)$.
Using Eqs.~(\ref{m3eqs}), we obtain
$d_1 = d_1 |r|^2 + 2 d_1 |t_1|^2 + (r_2 t_1^2 - r_1 t_2^2)$.
Hence, 
\begin{equation}
\label{triple}
r_2 t_1^2 - r_1 t_2^2 = d_1 \left(|t_2|^2 - |t_1|^2\right).
\end{equation}

\section{BIC condition}
In this section, we provide a detailed derivation of the BIC condition.
We start by solving system~(12) 
 from the main text of the article with respect to the exponents $\ee^{-2\ii \phi}$ and $\ee^{-2\ii \psi}$:
\begin{equation}
\label{eqmain1}
\begin{aligned}
	\ee^{-2\ii \phi} &= 
	\frac{q_c^2}{2 q_2 t_2^2}\cdot
	\left[2 \frac{q_1 q_2}{q_c^2} t_2 (-r_c t_1 + r_1 t_2) + \lt(r_2 t_1^2 - r_1 t_2^2 \rt) \pm \sqrt{\eta}\right]
	,\\
	\ee^{-2\ii \psi} &= 
	\frac{q_c^2}{2 q_1 t_1^2}\cdot
	\left[2 \frac{q_1 q_2}{q_c^2} t_1 (r_2 t_1 - r_c t_2) - \lt(r_2 t_1^2 - r_1 t_2^2 \rt) \pm\sqrt{\eta}\right].
\end{aligned}
\end{equation}
Here, the same signs of the square roots have to be used in both equations. The square root argument reads as
\begin{equation}
\label{eqeta}
\eta = \lt( r_2 t_1^2 - r_1 t_2^2 \rt)^2 + \frac{q_1 q_2}{q_c^2} \cdot 
4 t_1 t_2 (r_c t_1 - r_1 t_2) (r_c t_2-r_2 t_1).
\end{equation}

Equations~(\ref{eqmain1}) and~(\ref{eqeta}) can be substantially simplified taking into account the unitarity of matrices $\mat{S}_1$ and $\mat{S}_2$ (see the previous section).
Indeed, using Eqs.~(\ref{qprops}),~(\ref{m3eqs}), and~(\ref{triple}), we can rewrite Eqs.~(\ref{eqmain1}) as 
\begin{equation}
\label{eqmain2-app}
\begin{aligned}
	\ee^{-2\ii \phi} &= 
	d_1 \frac{q_c^2}{2 q_2 t_2^2}\cdot
	\lt[2\frac{|t_2|^2}{|q_c|^2}  - \lt( |t_1|^2 + |t_2|^2 \rt) \pm \sqrt{\xi}\rt],\\
	\ee^{-2\ii \psi} &= 
	d_1\frac{q_c^2}{2 q_1 t_1^2}\cdot
	\lt[2\frac{|t_1|^2}{|q_c|^2} - \lt( |t_1|^2 + |t_2|^2 \rt) \pm  \sqrt{\xi}\rt],
\end{aligned}
\end{equation}
where~$\xi = \eta/d^2$ is the following real number:
\begin{equation}
\label{eqxi-app}
\xi = \left(|t_1|^2+|t_2|^2\right)^2 - \frac{4 |t_1|^2 |t_2|^2}{|q_c|^2}.
\end{equation}

Let us prove that the exponent $\ee^{-2\ii \phi}$ defined by Eqs.~(\ref{eqmain2}) lies on the unit circle in the complex plane once $\xi \le 0$.
To do this, we first consider the value of the term inside the square brackets in the first equation of Eqs.~(\ref{eqmain2}).
When $\xi \le 0$, the first two terms inside the square brackets are real, whereas the last term containing the square root is an imaginary number.
Therefore, the squared modulus of the expression in the square brackets is calculated as 
$$
\lt[2\frac{|t_2|^2}{|q_c|^2}  - (|t_1|^2 + |t_2|^2)\rt]^2 -\xi =
4|t_2|^4 \cdot \frac{1-|q_c|^2}{|q_c|^4},
$$
This allows us to write the squared modulus of the exponent $\ee^{-2\ii \phi}$ defined by Eqs.~(\ref{eqmain2}) in the following form:
\begin{equation}
\label{eq001}
	\lt|\ee^{-2\ii \phi}\rt|^2 = 
	|d_1|^2 \frac{1-|q_c|^2}{|q_2|^2},
\end{equation}
Let us note that $|d_1| = 1$, since $d_1$ is the determinant of a unitary matrix $\mat{S}_1$.
Moreover, according to Eqs.~(\ref{m2simple}), the fraction in Eq.~(\ref{eq001}) equals unity.
Therefore, the right-hand side of Eq.~(\ref{eq001}) equals unity once $\xi \le 0$.
Similarly, one can prove that the same condition implies $|\ee^{-2\ii \psi}| = 1$.
Thus, the exponents in Eqs.~(\ref{eqmain1}) lie on a unit circle, and, therefore, the BICs exist once $\xi \le 0$. 

Let us note that once $|\ee^{-2\ii \phi}| = |\ee^{-2\ii \psi}| = 1$, we can calculate $\phi$ and $\psi$ by taking the argument of Eq.~(\ref{eqmain2}):
\begin{equation}
\label{phipsi}
\begin{aligned}
2 \phi 
&=
-\arg\frac{d_1 q_1}{t_2^2}
\pm\arctan\frac
{\sqrt{-\xi}}
{|t_1|^2+|t_2|^2 \left (1-2 |q_c|^{-2}\right)}
 + 2\pi m -\pi
\\
2 \psi 
&=
-\arg\frac{d_1 q_2}{t_1^2}
\pm\arctan\frac
{\sqrt{-\xi}}
{|t_2|^2+|t_1|^2\left (1-2 |q_c|^{-2}\right)}
 + 2\pi n -\pi
\end{aligned}
\end{equation}
Here, $m$ and $n$ are integers such that $\phi$ and $\psi$ are non-negative.
Moreover, since $n_\TE > n_\TM$, the inequality $\phi > \psi$ must hold.

It is important to note that the phases $\phi$ and $\psi$ in Eq~(\ref{phipsi}) depend on the coupling coefficients (elements of the scattering matrices $\mat{S}_1$ and $\mat{S}_2$).
These coefficients depend on the angle of incidence $\theta$. 
Therefore, Eqs.~(16) 
 from  the main text of the article do not define $w$ and $\theta$ in an explicit way.
However, the simple iteration method shows a good convergence rate here:
after estimating $w$ and $\theta$ using Eqs.~(16) 
(with the right-hand sides being calculated at some approximate value of $\theta$),
we calculate the right-hand sides again using the updated value of $\theta$.
After a few iterations, one obtains the position of the BIC with a high accuracy.

\section{Topological charge of the BIC}

Let us start by writing the five coupled-wave equations in the following form: 
\begin{equation}
\label{system}
\left\{
\begin{aligned}
R - t_1 U_2 - t_2 V_2  &= r I,\\
U_1 - r_1 U_2 - r_c V_2  &= t_1 I,\\
U_2 - r_c U_2 - r_2 V_2  &= t_2 I,\\
q_1 U_1 \ee^{\ii \phi} + q_c V_1 \ee^{\ii \psi} - U_2 \ee^{-\ii \phi} &= 0, \\
q_c U_1 \ee^{\ii \phi} + q_2 V_1 \ee^{\ii \psi} - V_2 \ee^{-\ii \psi} &= 0.
\end{aligned}
\right.
\end{equation}

If we assume $I = 0$, the system~(\ref{system}) becomes homogeneous with the non-trivial solution describing the field distribution without an incident wave, i.\,e. an eigenmode of the structure.
Taking into account Eqs.~(\ref{m2eqs}) and~(\ref{m3eqs}), we write this non-trivial solution in the following form:
\begin{equation}
\label{eqmode}
\begin{aligned}
R &= d_1 \ee^{\ii \phi} \lt( q_c t_1^* \ee^{\ii \psi} - q_1 t_2^* \ee^{\ii \phi}\rt) - t_2, \\
V_1 &= d_1 r_1^*-r_2  - d_1 q_1 (1 - r^*) \ee^{2\ii\phi}, \\
V_2 &= -1 + r + q_1 (r_1 - d_1 r_2^*) \ee^{2\ii\phi} + q_c (r_c + d_1 r_c^*) \ee^{\ii\phi}\ee^{\ii\psi}, \\
U_1 &=  d_1 q_c (1-r^*) \ee^{\ii\phi}\ee^{\ii\psi} - r_c - d_1 r_c^*, \\
U_2 &= -q_1 (r_c + d_1 r_c^*) \ee^{2\ii\phi} - q_c (r_2 - d_1 r_1^*) \ee^{\ii\phi}\ee^{\ii\psi}.
 \end{aligned}
\end{equation}

The BICs are non-radiating eigenmodes, hence, the amplitude $R$ in Eq.~(\ref{eqmode}) must be zero in the case of a BIC.
We denote this amplitude by 
\begin{equation}
\mathcal{P} = d_1 \ee^{\ii \phi} \lt( q_c t_1^* \ee^{\ii \psi} - q_1 t_2^* \ee^{\ii \phi}\rt) - t_2
\end{equation}
and use it to define the topological charge of the BIC in the main text of the article.

We have shown that at a BIC, the value of $\mathcal{P}$ is always zero.
Let us prove the opposite implication, i.\,e.\ that once $\mathcal{P}=0$, a BIC takes place.
To show this, we solve system~(\ref{system}) with respect to the amplitudes $V_1$, $V_2$, $U_1$, $U_2$, and $R$ assuming that the incident wave has unit amplitude ($I = 1$).
Taking into account Eqs.~(\ref{m2eqs}) and~(\ref{m3eqs}), we write the solution in the following form:
\begin{equation}
\label{solll}
\begin{gathered}
V_1 = -\ee^{-2 \ii \phi}\ee^{-2 \ii \psi}\frac{\mathcal P}{\mathcal D}, \;\; 
V_2 = - d_1 d_2 \frac{\mathcal P^*}{\mathcal D}, \;\; \\
U_1 = -\ee^{-2 \ii \phi}\ee^{-2 \ii \psi}\frac{\mathcal Q}{\mathcal D}, \;\; 
U_2 = - d_1 d_2\frac{\mathcal Q^*}{\mathcal D}, \\
R = r - d_1 d_2 t_1\frac{\mathcal Q^*}{\mathcal D} - d_1 d_2 t_2\frac{ \mathcal P^*}{\mathcal D}
= \ee^{2\ii\phi}\ee^{2\ii\psi} d_1 d_2\frac{\mathcal{D}^*}{\mathcal D}, 
\end{gathered}
\end{equation}
where $d_i = \det \mat{S}_i$ and 
\begin{equation}
\mathcal{Q} = d_1 \ee^{\ii \psi} \lt( q_c t_2^* \ee^{\ii \phi} - q_2 t_1^* \ee^{\ii \psi}\rt) - t_1.
\end{equation}

Assume that we are provided with real $\omega$, $w$, and $\theta$ such that $\mathcal{P} = 0$ .
In this case, two options are possible.
The first option is that the zeros of the numerators of $V_{1,2}$ in Eqs.~(\ref{solll}) are  canceled out by the zeros in the denominators, therefore, $\mathcal{D} = 0$ and a BIC takes place at the provided $(\omega, w, \theta)$ values.
If no BIC is supported by the structure at this $(\omega, w, \theta)$ point, we arrive to the second option: the whole fractions in $V_1$ and $V_2$ vanish.
In this case, as it is evident from the last two equations of system~(\ref{system}), both $U_1$ and $U_2$ are also zero.
This, according to the second and third equations of system~(\ref{system}), is possible only when $t_1=t_2=0$, which brings us to a contradiction, since $t_1$ and $t_2$ are elements of an arbitrary unitary scattering matrix $\mat{S_1}$ and are, in the general case, non-zero.

Therefore, we have shown that once $\mathcal{P} = 0$, the structure supports a BIC.
One can show that the same holds for the quantity $\mathcal{Q}$, therefore, not only $\mathcal{P}$ but also $\mathcal{Q}$ can be used to define the topological charge of a BIC.

\end{document}